\documentclass[12pt]{article}
\usepackage{graphicx}
\usepackage{amsmath}
\usepackage{feynmp}
\newcommand{\Z}{{\cal Z}}

\begin{document}

\begin{flushright}
HET--1161\\
TA--563
\end{flushright}

\begin{center}
{\large New Numerical Methods for Iterative or Perturbative Solution of Quantum
Field Theory\footnote{To appear in Proceedings of Fourth Workshop on Quantum
Chromodynamics, American University of Paris, 1--6 June 1998.}}\\[12pt]
Stephen C. Hahn\footnote{E-mail: {\tt hahn@het.brown.edu}.} and G. S. Guralnik\footnote{E-mail: {\tt gerry@het.brown.edu}.}\\
{\it Department of Physics, Brown University,}\\
{\it Providence RI 02912--1843}
\end{center}

\begin{abstract}
A new computational idea for continuum quantum Field theories is outlined. This
approach is based on the lattice source Galerkin methods developed by Garc\'\i
a, Guralnik and Lawson. The method has many promising features including
treating fermions on a relatively symmetric footing with bosons.  As a spinoff
of the technology developed for ``exact'' solutions, the numerical methods used
have a special case application to perturbation theory. We are in the process
of developing an entirely numerical approach to evaluating graphs to high
perturbative order.
\end{abstract}

\baselineskip=21pt

\section{General Approach}

Conventional numerical evaluation of quantum field theory involves
evaluating the path integral on a spacetime lattice using Mont\'e
Carlo integration methods. Mont\'e Carlo methods have been very
successful for many problems and have served to at least roughly
confirm the standard model of particle interactions.  However, many
serious issues are either not yet solved by this approach or not
likely to be calculable at all.  These include dealing with actions
that are not manifestly positive definite or which have important
effects from the details of fermionic interactions beyond the quenched
approximation.  Consequently, it is worthwhile to search for
supplementary methods of numerical calculation.

The source Galerkin method, although at a relatively primitive stage
of development, appears to work on both the lattice and the continuum
while dealing with fermions as easily as bosons.  Of course, the
continuum version of source Galerkin completely avoids the usual
lattice problem of fermionic multiplicity and the tricks used to
minimize errors from this problem.  Furthermore, our ``source
Galerkin'' method is less restrictive as to the class of allowed
actions.  Source Galerkin tends to use significantly less compute time
than Mont\'e Carlo methods but can consume significantly larger amounts
of memory.  The formulation of a source Galerkin approximation to a
given field theory can also require much more initial work and input
to the calculation than the usual Mont\'e Carlo approaches.

We confine this discussion to the continuum analysis of $\phi^4$
interactions; examples of lattice calculations have been given
elsewhere.\cite{group:initial,group:lawson:bosons,group:lawson:fermions}
Our continuum results to date include non-linear sigma
models and four-fermion interactions. We have gauge theory
calculations in progress.

We start with the differential equations satisfied by the the vacuum
function $\Z$ for a scalar field $\phi$ with interaction $g\phi^4/4$
coupled to a scalar source $J(x)$ which satisfies the equation:
\begin{equation}
\left( -\partial^2_D + M^2 \right) {\frac{\delta\Z}{\delta J(x)}}+
g\,{\frac{\delta^3\Z}{\delta J^3(x)}} -J(x) \Z = 0
\label{euc1}
\end{equation}
The source Galerkin technique is designed to directly solve functional
differential equations of this type. Before we proceed to outline a
solution technique, it is essential to point out that this equation by
itself does not uniquely specify a theory.\cite{group:ultralocal}
Since it is a set of third order differential equations at every point
in space time, it possesses, {\it a priori}, an infinite solution
set. The solution studied in this talk will be the usual one which
corresponds to the symmetry preserving solution obtained from
evaluating a path integral with real axis definitions for the regions
of integration. This is the solution that is regular in the coupling,
$g$, as it approaches zero.

If we write (\ref{euc1}) in the form:
\begin{equation}
\hat E_J \Z(J) = 0
\end{equation}
The source Galerkin method is defined by picking an approximation
$\Z^*(J)$ to the solution $\Z(J)$ such that
\begin{equation}
\hat E_J \Z^*(J) = R
\end{equation}
where $R$ is a residual dependent on $J$. We pick the parameters of
our approximation to make
this residual as small as possible on the average. To give this
statement a meaning, we define an inner product over the domain of
$J$: {\it i.e.\/} $(A,B)\equiv\int d\mu(J)\,A(J)B(J)$.  In addition, we
assume we have a collection of test functions which are members of a
complete set: $\{\varphi_i(J)\}$.  The source Galerkin minimization of
the residual $R$ is implemented by setting the parameters of our test
function $\Z^*(J)$ so that projections of test functions against the
residual vanish so that $|| \Z^* - \Z ||_2 \to 0$ as the number of
test functions $\to\infty$.

The equations defining the quantum field theory are differential
equations in the field sources and spacetime. We can deal with the 
continuum by taking advantage of our
knowledge of functional integration to evaluate integrals of the form:
\begin{equation}
I = \int [dJ] \exp\left[ -J^2(x)/\epsilon^2 \right] P(J)
\end{equation}
and use this to define an inner product of sources on the continuum
as follows:
\begin{equation}
(J(x_1)\cdots J(x_n), J(y_1)\cdots J(y_m))_J = \begin{cases}
\epsilon^{n+m}\delta_+\{x_1\cdots x_n y_1 \cdots y_m\}&n+m\ {\rm even}\\
0&{\rm otherwise}
\end{cases}
\end{equation}
where we have absorbed a factor of 2 by redefining $\epsilon$.
$\delta_+$ is defined by
\begin{gather}
\delta_+\{x\alpha\beta\cdots\} =
\delta(x-\alpha)\delta_+\{\beta\cdots\} +
\delta(x-\beta)\delta_+\{\alpha\cdots\} + \cdots,\\
\delta_+\{x\alpha\} = \delta( x-\alpha ).
\end{gather}

In addition to this inner product definition, we need good guesses for
approximate form for $\Z^*$ and numerical tools to calculate,
symbolically or numerically, various functions and their integrals,
derivatives, and so on.  For most of our calculations, we have found it
very useful to choose a lesser known class of functions, with
very suitable properties for numerical calculation, known as Sinc functions.
We take our notation for the Sinc functions from
Stenger:\cite{mono:stenger:sinc-methods} 
\begin{equation}
S(k,h)(x) = \frac{\sin( \pi(x-kh)/h)}{\pi(x-kh)/h}
\end{equation}
Sinc approximations satisfy many
identities,\cite{mono:stenger:sinc-methods} which make these functions
very easy to use for Galerkin methods, collocation, integration by
parts, and integral equations.

With the definition of a norm and set of expansion functions, we can
postulate an ans\"atz for $\Z$:
\begin{equation}
{\cal \Z}^* = \exp\left[\sum \int_{xy} J(x)G_2(x-y)J(y) + \cdots\right]
\end{equation}

This ans\"atz with only free field structure is very simple and one
would not expect to describe the entirety of a given theory with
it. However, the strength of the Galerkin method to force convergence
to correct answers is often strong enough to yield remarkably good
results for the lowest state of a theory.  Extremely good
approximations to theories are obtained as additional source structure
combined with the appropriate spacetime structure is added to the
ans\"atz.

In principle, it is possible to allow the functions multiplying the
sources to be totally arbitrary and then to fix them by expanding in
Sinc functions and apply the Galerkin technique.  The associated
computations soon demand more memory than is available from current
computing systems. Instead we can use our knowledge of the spectral
representations of field theory and graphical approaches (developed
from perturbation theory but we emphasize that our approach is not
perturbative) to introduce a beautiful and intuitive approach of
regulated Lehmann representations to produce candidates for
$\Z^*$. These build the appropriate spacetime Lorentz structure
into our approximations and make the operational cost of our numerical
approach independent of spacetime dimension.  Since any exact
two-point function can be represented as a sum over free two-point
functions an appropriate representation of the two-point function
greatly facilitates this approach.  We choose as the basis of our
numerical solutions a regulated Euclidean propagator structure:
\begin{align}
\label{green-function-form}
\Delta(m;x) \equiv
\int (dp) \frac{e^{ip\cdot x - p^2/\Lambda^2}}{p^2+m^2} &= \int (dp)
\int_0^\infty ds\, e^{ip\cdot x - p^2/\Lambda^2 - s(p^2+m^2)}\\ &=
\frac{1}{(2\pi)^d}\int_0^\infty
ds\,\left[\frac{\pi}{s+1/\Lambda^2}\right]^{d/2}e^{-sm^2 -
\frac{x^2}{4(s+1/\Lambda^2)}} 
\end{align}
This cutoff in the Galerkin test function as well as the definition of inner
product assures that we never have to address the issue of a
divergence during the course of a calculation, just as in the case of
lattice calculation where the lattice spacing serves as the cutoff.

This integral can be approximated using Sinc methods
\begin{equation}
\Delta(m;x) \approx \frac{h}{(2\pi)^d} \sum_{k=-N}^{N} \frac{1}{e^{kh}} \left[\frac{\pi}{z_k+1/\Lambda^2}\right]^{d/2}
\exp\left[-z_km^2 - \frac{x^2}{4(z_k+1/\Lambda^2)}\right],
\end{equation}
Here $z_k=e^{k*h}$ .
For practical purposes, a sum over eighty terms with $\Lambda^2=10$
gives accuracy out to twelve digits---more than adequate for most
hardware floating-point representations.  Thus we have a form for a two-point
scalar Green function, regulated by the scale $\Lambda^2$ with constant
computational cost regardless of spacetime dimension.  We can take derivatives
explicitly or by construction:
\begin{equation}
\partial^2 \Delta(m;x) = m^2\Delta(m;x) - \bar\delta(x)
\end{equation}
where $\bar\delta(x) = e^{-x^2\Lambda^2/4}$

From this representation, we can directly construct a fermion two-point
function:
\begin{equation}
S(m;x) = (\gamma\cdot\partial - m)\Delta(m;x)
\end{equation}
These representations mean that free scalar and free fermion results
are exact and immediate in any Galerkin evaluation of these trivial
cases. 
\section{Results: $\phi^4$} 
\begin{figure}
$$
\includegraphics[height=2in]{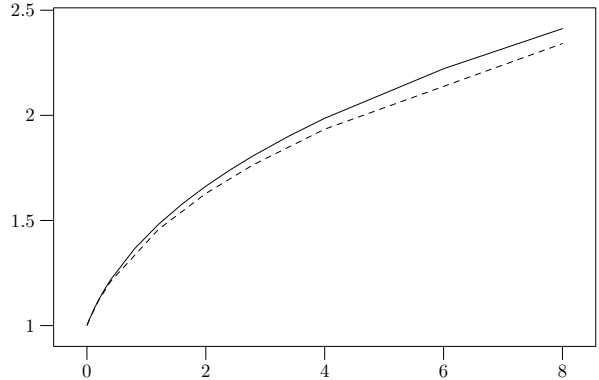}
$$
\caption{One dimensional $\phi^4$ mass gap versus coupling (dashed
line gives exact from Hioe and Montroll, 1975)}
\label{fig:lowest}
\end{figure}

We itemize some results obtained using a regulated single propagator with
parameters set by the Source Galerkin method.   At lowest order, our ans\"atz
for the generating functional is
\begin{equation}
\Z^* = \exp \int \frac{1}{2}j_xG_{xy}j_y.
\end{equation}
Results for this ans\"atz are given in Figure \ref{fig:lowest}.
These results are strikingly accurate and can matched up essentially exactly
with results of Mont\'e Carlo calculations in two and higher dimensions.

\begin{figure}
$$
\includegraphics[height=1in]{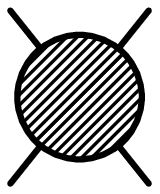}
$$
$$
{\includegraphics{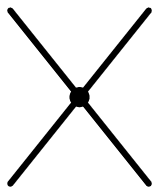}\includegraphics{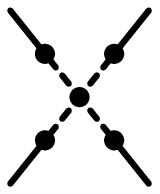}\includegraphics{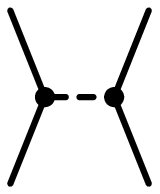}\includegraphics{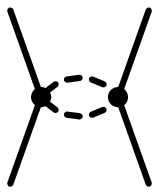}}
$$
\caption{Additional connector-based ans\"atzen for the four-point
function, $H$, in $\lambda\phi^4$.  In the bottom row, we have two
contact ans\"atzen on the left, followed by two mediated ans\"atzen.}
\label{figure:phi4-four-point}
\end{figure}

\begin{figure}
$$
\includegraphics[height=3in]{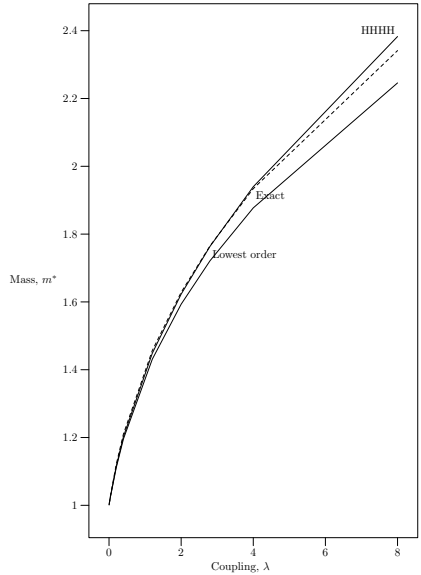}
$$
\caption{Comparison of four-$H$ approximation with lowest order and
exact answer, in one-dimension ($\Lambda^2=70$).} 
\label{p4:fig-hhhh-correction}
\end{figure}

We can enhance these results by including additional 4 source terms in
$\Z^*$. Some simple additional terms that we include with weights and
masses to be calculated using the Source Galerkin technique are the terms of the
forms given in Figure \ref{figure:phi4-four-point}.
The effect of adding a fourth order term is shown in
Figure \ref{p4:fig-hhhh-correction}.

\section{Perturbation Theory}

The representation that we have given for the free propagator
(\ref{green-function-form}) provides the basis for an entirely
numerical expansion of any Feynman graph.  We have examined the
possibility of of generating all graphs of a theory through automatic
functional differentiation of a vacuum amplitude and then numerically
analyzing the resultant expressions using Sinc
expansions.\cite{bsc:wang,group:perturbation} This method generates
and evaluates graphs at a speed which to our knowledge far exceeds any
method used in the past. We believe that it will be possible to
calculate to quite high orders in perturbation theory.

\section{Conclusions}

We have discussed a method for numerical calculations for field
theories on the continuum; this method being based on the original
lattice source Galerkin technique.\cite{group:initial} We presented a
Lorentz-invariant regulated representation derived from Lehmann
representations.  This approach has the computational advantages of
minimal memory utilization and parallelizable algorithms and also
allows direct representation of fermionic Green functions.  Finally, a
number of useful peripheral calculations can be made using this
approximate representation: one can calculate diagrams in a regulated
perturbation theory, as well as calculating dimensionally regularized
loops numerically.  In general, this technique of evaluating field
theories takes advantage of the symmetries of the Lorentz group;
future work includes the extension of the method to more general
internal groups, such as gauge groups or supersymmetry.

\section*{Acknowledgments} 

This work was supported in part by U. S. Department of Energy grant
DE-FG09-91-ER-40588---Task D.  The authors have been the beneficiaries
of many valuable conversations with R. Easther, W.-M. Wang,
S. Garc\'\i a, Z. Guralnik, J. Lawson, K. Platt, and P. Emirda\u g.
Certain results in this work were previously published in
Hahn.\cite{phd:hahn} Computational work in support of this research
was performed at the Theoretical Physics Computing Facility at Brown
University.


\end{document}